\documentclass[12pt,preprint]{aastex}

\usepackage{amsmath}
\usepackage{amssymb}
\usepackage{bm}

\shorttitle{Ion Acceleration Via Reconnection in Flares}
\shortauthors{Knizhnik et al.}

\begin{document}


\title{The Acceleration of Ions in Solar Flares During Magnetic
Reconnection}


\author{
K.~Knizhnik\altaffilmark{1}, 
M.~Swisdak\altaffilmark{2}, 
J.~F.~Drake\altaffilmark{2} }


\altaffiltext{1}{Department of Physics and Astronomy, Johns Hopkins
University, Baltimore, MD 21218; kknizhni@pha.jhu.edu}
\altaffiltext{2}{University of Maryland, College Park, MD 20742;
swisdak@umd.edu, drake@umd.edu}


\begin{abstract}

The acceleration of solar flare ions during magnetic reconnection is
explored via particle-in-cell simulations that self-consistently and
simultaneously follow the motions of both protons and $\alpha$
particles.  We show that the dominant heating of thermal ions during
guide field reconnection, the usual type in the solar corona, results
from pickup behavior during the entry into reconnection exhausts.  In
contrast to anti-parallel reconnection, the temperature increment is
dominantly transverse, rather than parallel, to the local magnetic
field. A comparison of protons and alphas reveals a mass-to-charge
($M/Q$) threshold in pickup behavior that favors heating of high-$M/Q$
ions over protons, which is consistent with impulsive flare
observations.

\end{abstract}




\section{INTRODUCTION}\label{intro}

The generation of energetic particles during flares remains a central
unsolved issue in solar physics.  Extensive observational evidence
indicates that a substantial fraction of the energy released during a
flare rapidly accelerates charged particles, with electrons reaching
$\mathcal{O}(1)$ MeV and ions $\mathcal{O}(1)$ GeV/nucleon
\citep{emslie04a}.  Explaining this energization requires accounting
not only for the relevant energy and time scales but also the
resulting spectra, which exhibit a common shape for almost all ion
species.  At the same time, high mass-to-charge ($M/Q$) ions are
greatly over-represented in flares, with abundances as much as two
orders of magnitude higher than normal coronal values
\citep{mason94a,mason07a}.

In impulsive flare models, magnetic reconnection is the ultimate
energy source, and so it is natural to consider theories in which
reconnection also plays a role in particle acceleration.  Some models,
including those that rely on interactions with magnetohydrodynamic
(MHD) waves \citep{miller98a,petrosian04a} or shock acceleration
\citep{ellison85a,somov97a}, use reconnection only as an indirect
source that provides an environment in which other energization
processes can occur.  In other models, the magnetic energy released
during reconnection is more directly channeled to particles through
various processes --- DC electric fields \citep{holman85a},
interactions with multiple magnetic islands \citep{onofri06a},
first-order Fermi acceleration \citep{drake10a}, or the pickup of
collisionally ionized neutrals \citep{wu96a}.

Also a member of this latter group is the direct heating of ions in
reconnection exhausts \citep{kraussvarban06a,drake09a}, in which the
perpendicular and parallel (relative to the magnetic field)
temperatures of ions jump after traversing the narrow boundary layer
separating the ions that flow slowly in from upstream from the
reconnection exhaust, which travels at the Alfv\'en speed
$B/\sqrt{4\pi\rho}$ where $B$ is the strength of the magnetic field
and $\rho$ is the density.  However these works considered the weak
guide field\footnote{A guide field is a component of the magnetic
field perpendicular to the reconnection plane.  Most coronal
reconnection is guide field reconnection.} limit in which ion heating
is parallel, rather than transverse, to the local magnetic field;
\cite{cranmer03a} have shown that in the extended solar corona,
$T_{\perp} \gg T_{\parallel}$.  Subsequently, \cite{drake09b} used
test particles\footnote{Test particles are particles that move under
the influence of the simulation's electromagnetic fields, but do have
any self-consistent effect on the computation.} in a Hall MHD
simulation with a large guide field (five times larger than the
reconnecting field) to confirm that ions above a critical value of
$M/Q$ become demagnetized.  They suggested that ions crossing into
reconnection outflows can become non-adiabatic, and hence behave like
pickup particles\footnote{The pickup process refers to the ionization
of a neutral atom with velocity $\approx 0$ embedded in a
high-velocity plasma. It plays an important role throughout the
heliosphere, and particularly in the solar wind.}  \citep{mobius85a},
while gaining an effective thermal velocity equal to the Alfv\'en
speed and derived a $M/Q$-based threshold for this behavior.  This
process is similar to an earlier proposal by \cite{wu96a} that ion
acceleration in impulsive flares can occur via reconnection-associated
pickup, although in that case the accelerated ions were produced by
neutral-particle ionization in the lower corona.  Later hybrid
simulations by \cite{wang01a} confirmed that injected protons
(mimicking newly ionized particles) did behave like pickup particles
in this scenario.

In this Letter, we use a kinetic particle-in-cell (PIC) simulation to
track two types of ions self-consistently, i.e. without resorting to
test particles, to determine whether particles above the critical
value of $M/Q$ behave like pickup particles in dynamic electromagnetic
fields. Ions with $M/Q$ below the threshold derived in \cite{drake09b}
(protons, in this case) are adiabatic and undergo very little heating
as they move between the upstream plasma and the reconnection exhaust,
while particles above the threshold ($\alpha$ particles) gain an
effective thermal velocity equal to the exhaust velocity after
crossing the narrow boundary layer surrounding the exhaust.  

The transition between adiabatic and non-adiabatic behavior depends on
the ratio between a particle's cyclotron period and the the time it
takes to cross the boundary layer \citep{drake09b}. An adiabatic
particle turns sharply in the outflowing direction upon entering the
exhaust, conserving its magnetic moment $\mu=m\delta
v_{\perp}^{2}/2B$, where $\delta v_{\perp}$ is the ion perpendicular
velocity with the $\mathbf{E} \boldsymbol{\times} \mathbf{B}$
contribution subtracted (the ion perpendicular velocity is taken
relative to $\mathbf{B}$).  However, particles which behave
non-adiabatically move in the direction of the local electric field
upon entering the exhaust and not in the direction of the local
$\mathbf{E} \boldsymbol{\times} \mathbf{B}$ velocity. The sudden
change from slow upstream inflow to downstream Alfv\'enic outflow
causes particles with high $M/Q$ to see a jump in their magnetic
moments.

\section{NUMERICAL SIMULATIONS}
\label{simulations}

We carry out simulations using the code {\tt p3d} \citep{zeiler02a}.
Like all PIC codes, it tracks individual particles ($\approx 10^9$ in
this work) as they move through electromagnetic fields that are
defined on a mesh.  Unlike more traditional fluid representations
(e.g., MHD), PIC codes correctly treat small lengthscales and fast
timescales, which are particularly important for understanding the
x-line and separatrices during magnetic reconnection.

The simulated system is periodic in the $x-y$ plane, where flow into
and away from the x-line are parallel to $\mathbf{\hat{y}}$ and
$\mathbf{\hat{x}}$, respectively, and the guide magnetic field and
reconnection electric field parallel $\mathbf{\hat{z}}$.  The initial
magnetic field and density profiles are based on the Harris
equilibrium \citep{harris62a}.  The reconnecting magnetic field is
given by $B_x=\tanh[(y-L_y/4)/w_0]- \tanh[(y-3L_y/4)/w_0]-1$, where
$w_0 $ and $L_y$ are the half-width of the current sheets and the box
size in the $\mathbf{\hat{y}}$ direction.  The density comprises an
ambient background and two current sheets in which the density rises
in order to maintain pressure balance with the magnetic field.  We
initiate reconnection with a small initial magnetic perturbation that
produces a single magnetic island on each current layer.

The code is written in normalized units in which magnetic fields are
scaled to the asymptotic value of the reversed field $B_{0x}$,
densities to the value at the center of the reconnecting current sheet
minus the uniform background density, velocities to the proton
Alfv\'en speed $c_{A}=B_{0x}/\sqrt{4\pi m_pn_0}$, times to the inverse
proton cyclotron frequency in $B_{0x}$,
$\Omega_{px}^{-1}=m_pc/eB_{0x}$, lengths to the proton inertial length
$d_p=c_{A}/\Omega_{px}$ and temperatures to $m_p c_{A}^2$.

The proton to electron mass ratio is taken to be $25$, in order to
minimize the difference between pertinent length scales and hence run
as large a simulation domain as possible.  It has been shown
\citep{shay98a,hesse99a,shay07a} that the rate of magnetic
reconnection and structure of the outflow exhaust do not depend on
this ratio, and neither, therefore, does the ion heating examined
here, which depends only on the exhaust geometry. The simulation
assumes $\partial/\partial z=0$, i.e. that field and particle
quantities do not vary in the out-of-plane direction, making this a
two-dimensional simulation.

In addition to the usual protons and electrons, we also include a
number density of $1\%$ $^4$He$^{++}$ ($\alpha$) particles in the
background particle population and gave them an initial temperature
equal to that of the protons.  This number density does not affect the
reconnection dynamics appreciably, while still providing a large
sample of particles with $M/Q > 1$, where $M/Q$ is normalized to the
proton value. Each particle (protons and $\alpha$'s) is assigned a
unique tag number, allowing individual particles to be tracked
throughout the simulation.

\begin{figure}
\epsscale{1.70}
\includegraphics[keepaspectratio,width=\columnwidth]{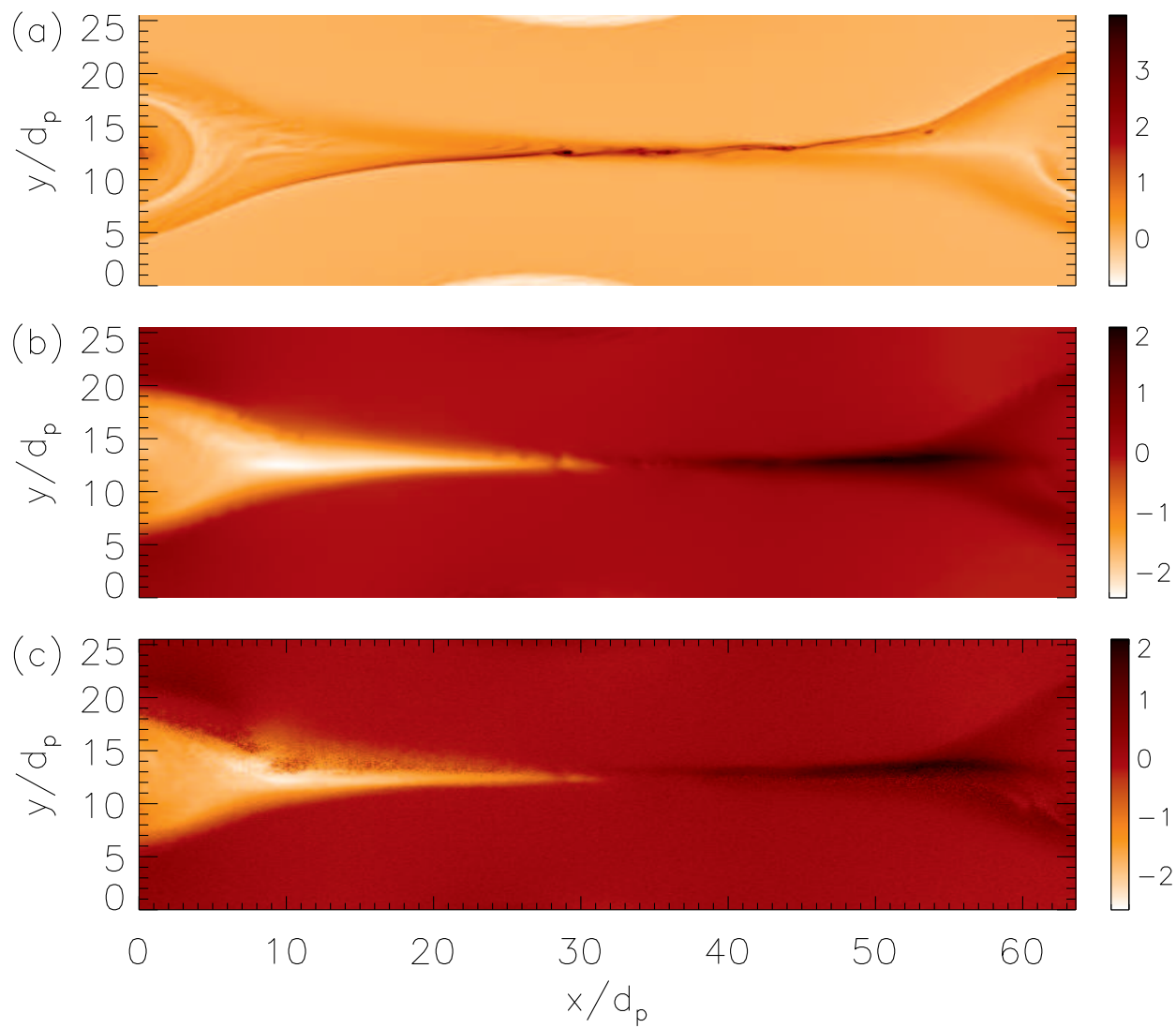}
\caption{\label{JandVX} Overview of a PIC simulation with an initial
guide field $B_{0z}=2B_{0x}$.  Panel (a): the total out-of-plane
current density $J_{z}$; panel (b): the proton outflow velocity
$v_{px}$; panel (c): the $\alpha$ particle outflow velocity $v_{\alpha
x}$}
\end{figure}

In Fig.~\ref{JandVX} we show an overview of results from a simulation
with a computational domain $L_x\times L_y=102.4 \times 51.2\,d_p$ and
an initial guide field $B_{0z} = 2.0B_{0x}$ at
$t=200\Omega_{px}^{-1}$. The grid spacing for this run is
$0.025\,d_p$, the electron, proton, and $\alpha$ temperatures,
$T_e=T_p=T_{\alpha}=0.25m_pc_{A}^2$, are initially uniform and the
velocity of light is $15c_{A}$. The half-width of the initial current
sheet, $w_0$, is $1\,d_p$ and the background density is $0.2$.  Panel
(a) depicts the total out-of-plane current density $J_{z}$ centered
around the x-line (at $x/d_p \approx 32$ and $y/d_p \approx 13$) of
one of the current sheets.  Magnetic field lines (not shown) roughly
trace contours of $J_z$.

Ambient plasma from above and below the current sheet slowly flows
toward the current sheet while embedded in oppositely directed
magnetic field (to the right above the layer, to the left below).
Reconnected field lines are highly bent and, to reduce their magnetic
tension, rapidly move away from the x-line, dragging plasma with them.
Panels (b) and (c) show the proton and $\alpha$ outflow velocities
$v_{px}$ and $v_{\alpha x}$.  The similarity between the two makes it
clear that both the protons and the $\alpha$'s participate in the
reconnection outflow which, outside of the immediate vicinity of the
X-line, has a magnitude of $v_{x}\sim cE_y/B_z\sim c_A$ ($\approx 1$,
in our normalized units).  A comparison of Fig.~\ref{JandVX} to frames
(a) and (b) of Fig.~1 in \cite{drake09b} (which shows results from a
run otherwise identical but for the presence of the $\alpha$
particles) demonstrates that the $\alpha$'s do not significantly
change the structure of the reconnection exhaust.

\section{ION PICKUP AND HEATING}\label{pickup}

\begin{figure}
\includegraphics[keepaspectratio,width=0.5\columnwidth]{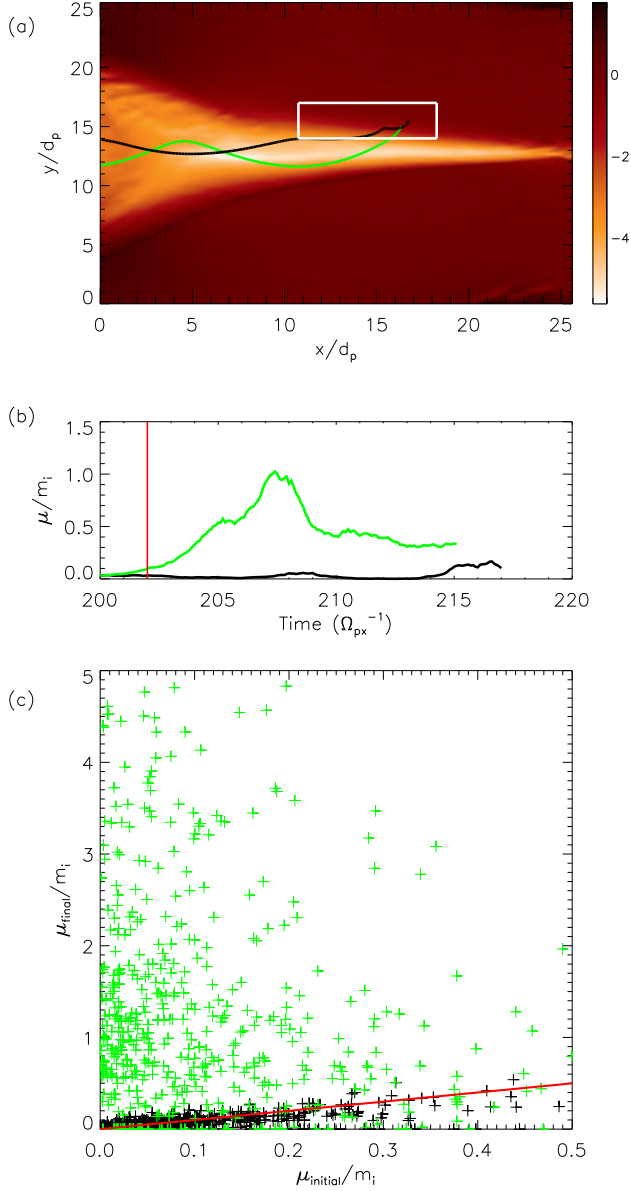}
\caption{\label{mudata} Panel (a): Trajectories for a proton (black) and
$\alpha$ particle (green) randomly picked from a $7.5 \times 3\,d_p$
box (shown in white) are overlaid on a snapshot of $E_y$.  Panel (b):
The magnetic moment per mass as a function of time for the two
particles in (a). The red line represents the time of the snapshot of
$E_y$. Panel (c): For 500 protons (black) and 500 $\alpha$'s (green)
selected at random from the white box in (a), their final magnetic
moments plotted against their initial magnetic moments. The red line
has unit slope and represents the expected result if $\mu/m$ were
invariant.}
\end{figure}

Particle acceleration is controlled by the structure and magnitude of
the electric field. During reconnection a strong transverse electric
field $E_y = -E_zB_{0z}/B_y$ develops in the exhaust to force
$\mathbf{E} \cdot \mathbf{B}=0$; its structure to the left of the
x-line is shown in the background of Fig.~\ref{mudata}(a).  Particles
enter the exhaust with a velocity $v_y \sim 0.1 c_A$. Any energy gain
is determined by whether particles crossing the exhaust boundary,
which has scale length given by the ion sound Larmor radius $\rho_s$
($=v_s/\Omega_p$, where $v_s$ is the plasma sound speed), are
adiabatic. Non-adiabatic particles cross the boundary in a time
$\tau_c$ that is short compared with their cyclotron period, $\tau_c
\approx \rho_s/v_y \approx 10\rho_s/c_A < \pi/\Omega_{px}$, or
\begin{equation}\label{mqcrit}
\frac{M}{Q}>\left(\frac{5\sqrt{2}}{\pi}\right)\sqrt{\beta_{px}}
\end{equation} where $\beta_{px}=8\pi nT/B_{0x}^2$ \citep{drake09b}.
Thus, in the present simulations, where the upstream $\beta_{px}=0.2$,
equation \ref{mqcrit} gives $M/Q > 1$ for non-adiabatic behavior and
so protons are marginally adiabatic while $\alpha$'s ($M/Q = 2$) are
not.  Since $E_y < 0$ in this case, non-adiabatic positively charged
ions entering the exhaust from below will be pushed out immediately,
preventing them from being caught up in the exhaust. However,
non-adiabatic positively charged ions entering from the top will find
themselves essentially at rest in the simulation frame while the
outflow moves past at roughly the Alfv\'en speed.  Such particles will
undergo an $\mathbf{E} \boldsymbol{\times} \mathbf{B}$ drift, but with
a ``thermal velocity'' equal to the Alfv\'en speed and have
trajectories resembling cycloids.  This process is analogous to that
undergone by stationary neutral atoms surrounded by the moving solar
wind.  If ionized, the new ion first moves in the direction of the
motional electric field in order to gain the necessary energy to flow
with the rest of the wind.  As it gets ``picked up'', it gains a
thermal velocity equal to the solar wind velocity \citep{mobius85a}.

We randomly selected 500 protons and 500 $\alpha$ particles from the
$7.5 \, d_p \times 3 \, d_p$ box upstream of the exhaust shown in
Fig.~\ref{mudata}(a) at $t=200\Omega_{px}^{-1}$ and followed their
trajectories for 25 $\Omega_{px}^{-1}$.  In Fig.~\ref{mudata}(a) we
plot a representative trajectory for a proton, shown in black, and an
$\alpha$ shown in green, over a background of $E_y$. (Note that the
overlaid trajectories in (a) are calculated in the fully
self-consistent simulation, while the background of $E_y$ is a
snapshot from $t=202 \, \Omega_{px}^{-1}$.) The proton, which remains
adiabatic, immediately moves downstream upon entering the exhaust,
while the $\alpha$ particle moves in the direction of $E_y$ before
being picked up by the $\mathbf{E} \boldsymbol{\times} \mathbf{B}$
drift.  Panel (b) displays the time evolution of the proton (black)
and $\alpha$ (green) magnetic moments (scaled by mass).  The vertical
red line corresponds to the time at which $E_y$ is shown in (a). After
crossing the boundary layer into the exhaust, the $\alpha$ becomes
demagnetized, as indicated by the jump in $\mu$, a trend seen for all
of the tracked $\alpha$'s.  In (c) we plot the magnetic moments of all
500 protons (black) and all 500 $\alpha$'s (green) after entering the
exhaust versus their moments at $t=200\Omega_{px}^{-1}$.  For each
particle $\mu_{\text{final}}$ was measured when the particle crossed a
specified horizontal position at the downstream edge of the exhaust,
around $6 \, d_p$ in Fig.~\ref{mudata}(a).  For reference, we overplot
a line of unit slope, which corresponds to exact $\mu$ conservation.
The clustering of protons near this line and large values of $\mu/m$
reached by the $\alpha$'s clearly shows the adiabatic nature of the
former and the non-adiabatic nature of the latter.

\begin{figure}
\includegraphics[keepaspectratio,width=0.66\columnwidth]{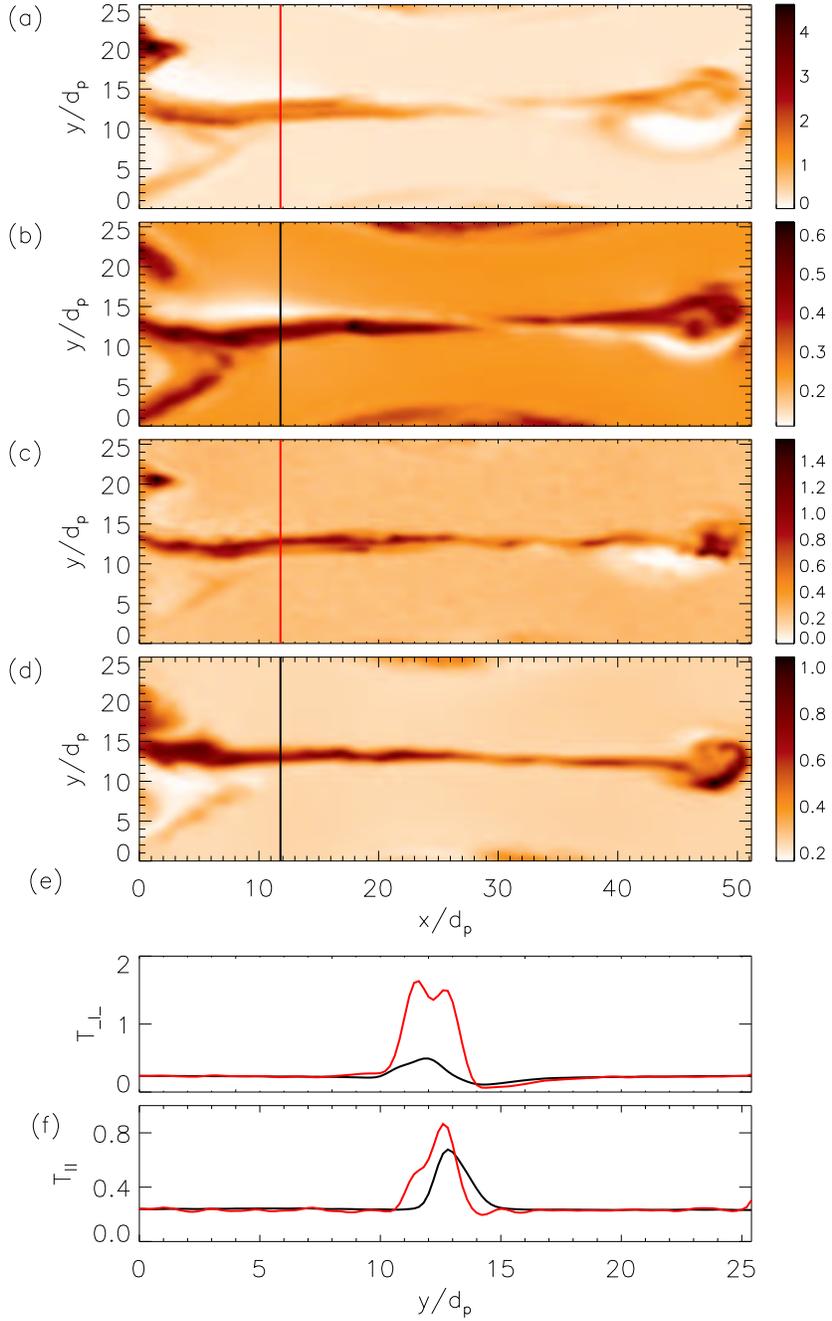}
\caption{\label{temperature} Spatially smoothed temperature
components at $t=200 \Omega_{px}^{-1}$.  Panel (a): $T_{\perp}$ for
$\alpha$ particles; panel (b): $T_{\perp}$ for protons; panel (c):
$T_{\parallel}$ for $\alpha$'s; panel (d): $T_{\parallel}$ for
protons. Panels (e) and (f): Cuts through the exhaust of the
perpendicular and parallel components, respectively, for $\alpha$'s
(red) and protons (black).  The locations of the cuts are shown by the
vertical lines in (a)-(d).}
\end{figure}

In Fig.~\ref{temperature} we show the temperatures of both
species. Panels (a) and (b) depict the perpendicular (to the magnetic
field) $\alpha$ and proton temperatures, while (c) and (d) show the
parallel temperatures. The $\alpha$ temperature increase is greater
than the proton temperature in the perpendicular direction (note the
different color bar scales in the two panels).  Indeed, the
temperature increase of the $\alpha$'s is more than mass proportional,
consistent with observations \citep{cranmer03a}.  This is also evident
in frames (e) and (f), which are cuts through the perpendicular and
parallel temperature plots for the $\alpha$'s (red) and protons
(black). The weak heating of the protons is consistent with the
adiabatic behavior shown in Fig.~\ref{mudata}.  The analysis of
\cite{drake09b} predicts that, with a guide field of $B_{0z} =
2B_{0x}$, the proton temperature will change by
\begin{equation} \label{protondt}
\Delta T_{\parallel} =\frac{B_{0x}^2}{B_{0z}^2}v_x^2 \sim
\frac{v_x^2}{4}\text{;}
\qquad \Delta T_{\perp}=0
\end{equation} 
in the exhaust, and the $\alpha$ temperature will change by:
\begin{equation} \label{alphadt}
\Delta T_{\parallel}=0\text{;} \qquad \Delta
T_{\perp}=\frac{1}{2}m_\alpha v_x^2
\sim 2v_x^2.
\end{equation} 
For $v_x^2 \sim 2$ (see Fig.~\ref{JandVX}) these jumps are in
reasonable agreement with the observed variations.  Differences from
the predicted values, in particular the changes in $T_{\perp}$ for the
protons and $T_{\parallel}$ for the $\alpha$'s, presumably arise from
corrections to equations (\ref{protondt}) and (\ref{alphadt}) due to a
mixture of adiabatic and non-adiabatic behavior by the particles.

\begin{figure}
\includegraphics[keepaspectratio,width=\columnwidth]{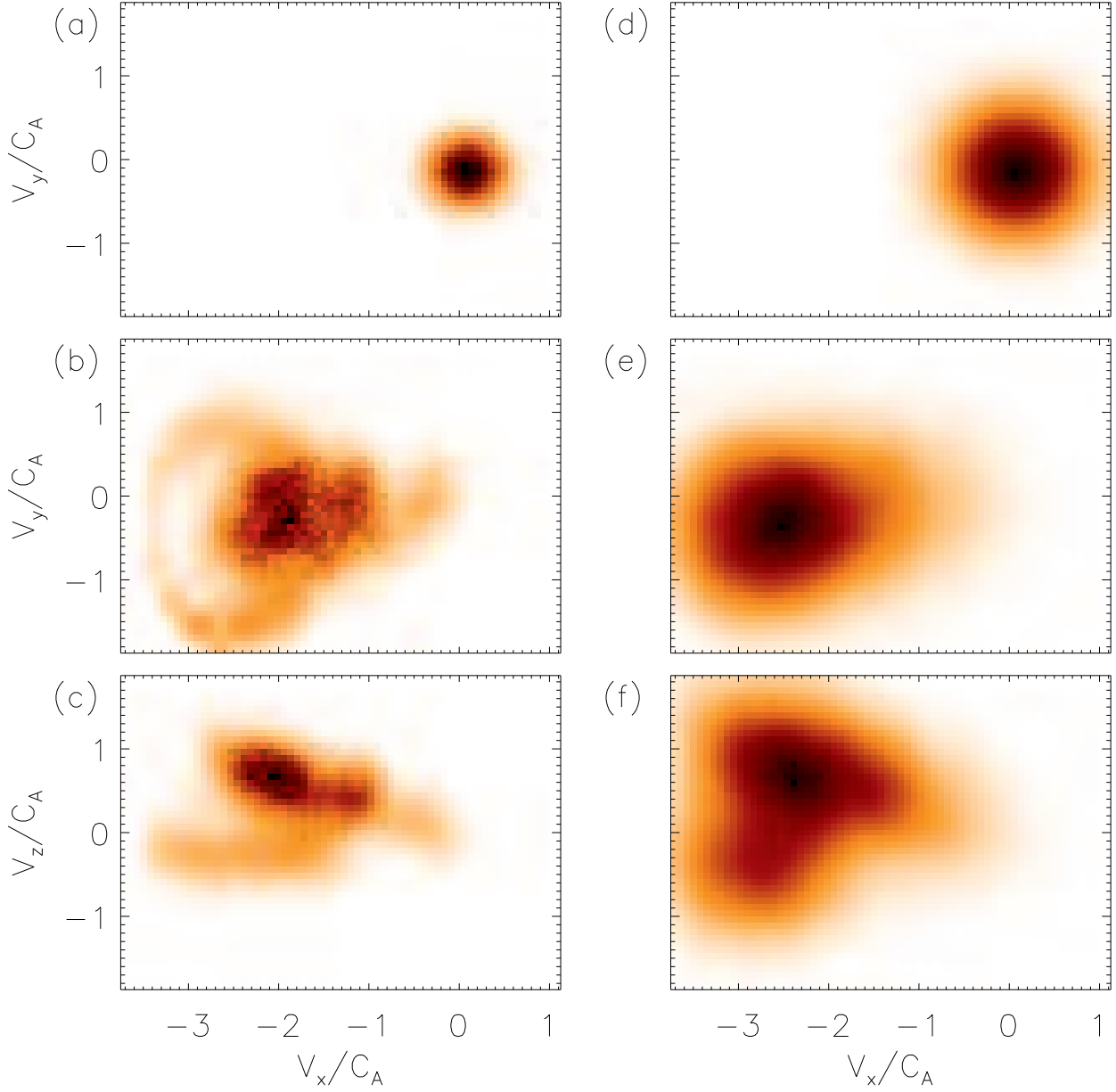}
\caption{\label{vel2d} 2D velocity distribution functions upstream and
downstream of the exhaust.  Panels (a) and (d): upstream $v_x-v_y$
distributions for $\alpha$'s and protons, respectively.  Panels (b)
and (e): downstream $v_x-v_y$ distributions for $\alpha$'s and
protons; panels (c) and (f): downstream $v_x-v_z$ distributions. }
\end{figure}

In Fig.~\ref{vel2d}, we show how the velocity distributions of the
$\alpha$'s and protons change as they move from the upstream to the
downstream region. Panels (a) and (d) depict the upstream $\alpha$ and
proton velocity distribution in the $v_x-v_y$ plane. The protons and
$\alpha$'s were given the same initial temperature, so upstream of the
exhaust, the protons' mean thermal velocity is higher than the
$\alpha$ particles'.  The small negative $v_{y}$ component upstream in
both species shows the inflow toward the reconnection exhaust.  After
crossing the narrow boundary layer, the $\alpha$'s get picked up by
the Alfv\'enic outflow (at this time the local density is $\sim$ 0.2,
so the local Alfv\'en speed is $\sim \sqrt{5}$), and their thermal
velocity increases much more than that of the adiabatic protons.  The
downstream velocity distributions for the $\alpha$'s and protons are
shown in the $v_x-v_y$ plane in (b) and (e) and in the $v_x-v_z$ plane
in (c) and (f), both calculated inside a box located between $5-12.5\,
d_p$ in the x-direction and $11.4-13.4\, d_p$ in the y-direction.
Since the dominant $\mathbf{B}$ component is the guide field, $v_x$
and $v_y$ are essentially perpendicular velocity components, and $v_z$
the parallel velocity.  The protons exhibit very little heating in the
$v_x - v_y$ plane (Fig.~\ref{vel2d}(e)), consistent with adiabatic
behavior. They are modestly heated in the z-direction
(Fig.~\ref{vel2d}(f)), consistent with $T_{\parallel}$ in
Eq.~\ref{protondt}. The $\alpha$ particles are strongly heated in the
y-direction (Fig.~\ref{vel2d}(b)) and are beginning to form a ring
distribution that is characteristic of pickup behavior. There is
modest heating of $\alpha$ particles in the z-direction
(Fig.~\ref{vel2d}(c)), but the similar structure in (c) and (f)
suggests that the heating mechanism is the same for both protons and
$\alpha$'s, and it is possible that the $\alpha$ particles are not
completely non-adiabatic.

\section{DISCUSSION}

Using self-consistent tracking of particle trajectories, we have shown
that ions above a critical mass-to-charge threshold \citep{drake09b}
behave like pickup ions in reconnection exhausts, with $\mu$ changing
due to a sharp increase in $v_{\perp}$.  Ion energy increments of
$\approx 25$ keV/nucleon are predicted for typical coronal parameters
of $B = 50$ G and $n = 10^9/\text{cm}^3$.  Ions below this threshold
are only weakly heated.  This transition only exists for reconnection
with a guide field which, however, is the typical case in the solar
corona.  Coronal observations have revealed that the abundances of
high mass-to-charge ions are enhanced in solar flares, with the
strength of the enhancement depending only on $M/Q$. The fact that we
observe non-adiabatic behavior and associated strong heating for
particles with $M/Q > 1$, while the proton heating remains weak,
suggests that reconnection might explain the abundance enhancements in
impulsive flares.  Abundance enhancements should occur because high
$M/Q$ ions are heated at lower values of the reconnecting magnetic
field strength (see Eq.~\ref{mqcrit}) than protons. Furthermore, the
increase in $T_{\perp}/T_{\parallel}$ in the exhaust seen here is
consistent with that observed in the extended solar corona
\citep{cranmer03a}, although it should be noted that the number
density of $\alpha$ particles used here is slightly less than what is
observed in the corona ($n_{\alpha\text{,corona}} \sim 5\%-10\%$).

Observations near 1 AU of solar wind reconnection events with the Wind
spacecraft (as, for example, in \cite{phan10a}) should be able to
measure $T_{\parallel}$ and $T_{\perp}$ for both protons and $\alpha$
particles in order to test the mechanism suggested in this work.
Moreover, provided its instrumentation can differentiate between
different $M/Q$ ions, the upcoming Solar Probe Plus mission, with a
planned perihelion of $\approx 9 R_{\odot}$ (which lies within the
outer corona), should also provide an excellent test of these
predictions.

Finally, it is widely believed that some process converts a fraction
of the energy found in the convective motions of the solar photosphere
into the heat that ensures the continuous existence of a
$\mathcal{O}(10^6\text{ K})$ corona and accelerates the solar wind.
Broadly speaking the two most likely candidates are wave heating ---
in which oscillations generated in the photosphere travel into the
corona, develop into turbulence, and eventually dissipate --- and
magnetic reconnection, in which the topological reorganization of the
magnetic field releases energy and heats the plasma.

Measurements by the Solar Ultraviolet Measurements of Emitted
Radiation (SUMER) and Ultraviolet Coronal Spectrometer (UVCS)
instruments of the SOHO (Solar and Heliospheric Observatory)
spacecraft provide significant constraints on any theory of coronal
heating.  In particular, at heights of $2-3 R_{\odot}$ protons have a
slight temperature anisotropy (in the $T_{\perp} > T_{\parallel}$
sense) while heavier ions (represented by $O^{5+}$) are strongly
anisotropic, with $T_{\perp}/T_{\parallel} \gtrsim 10$
\citep{cranmer03a}.  Interestingly, the process discussed in this work
should be active in the region in question and produces temperature
anisotropies consistent with these results.  

\acknowledgments

This work has been supported by an NSF Grant ATM-0903964 and NASA
grants APL-975268 and NNN06AA01C.  Computations were carried out at
the National Energy Research Scientific Computing Center.


\begin{thebibliography}{23}
\expandafter\ifx\csname natexlab\endcsname\relax\def\natexlab#1{#1}\fi

\bibitem[{Cranmer \& van Ballegooijen(2003)}]{cranmer03a}
Cranmer, S.~R., \& van Ballegooijen, A.~A. 2003, Ap. J., 594, 573

\bibitem[{Drake {et~al.}(2009{\natexlab{a}})Drake, Cassak, Shay, Swisdak, \&
  Quataert}]{drake09b}
Drake, J.~F., Cassak, P.~A., Shay, M.~A., Swisdak, M., \& Quataert, E.
  2009{\natexlab{a}}, Ap. J., 700, L16

\bibitem[{Drake {et~al.}(2010)Drake, Opher, Swisdak, \& Chamoun}]{drake10a}
Drake, J.~F., Opher, M., Swisdak, M., \& Chamoun, J.~N. 2010, Ap. J., 709, 963

\bibitem[{Drake {et~al.}(2009{\natexlab{b}})Drake, Swisdak, Phan, Cassak, Shay,
  Lepri, Lon, Quataert, \& Zurbuchen}]{drake09a}
Drake, J.~F., {et~al.} 2009{\natexlab{b}}, J. Geophys. Res., 114

\bibitem[{Ellison \& Ramaty(1985)}]{ellison85a}
Ellison, D.~C., \& Ramaty, R. 1985, Ap. J., 298

\bibitem[{Emslie {et~al.}(2004)Emslie, Kucharek, Dennis, Gopalswamy, Holman,
  Share, Vourlidas, Forbes, Gallagher, Mason, Metcalf, Mewaldt, Murphy,
  Schwartz, \& Zurbuchen}]{emslie04a}
Emslie, A.~G., {et~al.} 2004, J. Geophys. Res., 109

\bibitem[{Harris(1962)}]{harris62a}
Harris, E.~G. 1962, Nuovo Cim., 23, 115

\bibitem[{Hesse {et~al.}(1999)Hesse, Schindler, Birn, \& Kuznetsova}]{hesse99a}
Hesse, M., Schindler, K., Birn, J., \& Kuznetsova, M. 1999, Phys. Plasmas, 6,
  1781

\bibitem[{Holman(1985)}]{holman85a}
Holman, G.~D. 1985, Ap. J., 293, 584

\bibitem[{Krauss-Varban \& Welsch(2006)}]{kraussvarban06a}
Krauss-Varban, D., \& Welsch, B.~T. 2006, Proceedings of the International
  Astronomical Union, 2, 89

\bibitem[{Mason(2007)}]{mason07a}
Mason, G.~M. 2007, Space Sci. Rev., 130, 231

\bibitem[{Mason {et~al.}(1994)Mason, Mazur, \& Hamilton}]{mason94a}
Mason, G.~M., Mazur, J.~E., \& Hamilton, D.~C. 1994, Ap. J., 425, 843

\bibitem[{Miller(1998)}]{miller98a}
Miller, J.~A. 1998, Space Sci. Rev., 86, 79

\bibitem[{M{\"{o}}bius {et~al.}(1985)M{\"{o}}bius, Hovestadt, Klecker, Scholer,
  Gloeckler, \& Ipavich}]{mobius85a}
M{\"{o}}bius, E., Hovestadt, D., Klecker, B., Scholer, M., Gloeckler, G., \&
  Ipavich, F.~M. 1985, Nature, 318, 426

\bibitem[{Onofri {et~al.}(2006)Onofri, Isliker, \& Vlahos}]{onofri06a}
Onofri, M., Isliker, H., \& Vlahos, L. 2006, Phys. Rev. Lett., 96

\bibitem[{Petrosian \& Liu(2004)}]{petrosian04a}
Petrosian, V., \& Liu, S. 2004, Ap. J., 610, 550

\bibitem[{Phan {et~al.}(2010)Phan, Gosling, Paschmann, Pasma, Drake,
  {\O}ieroset, Larson, Lin, \& Davis}]{phan10a}
Phan, T.~D., {et~al.} 2010, Ap. J.. Lett., 719, L199

\bibitem[{Shay {et~al.}(1998)Shay, Drake, Denton, \& Biskamp}]{shay98a}
Shay, M.~A., Drake, J.~F., Denton, R.~E., \& Biskamp, D. 1998, J. Geophys.
  Res., 103, 9165

\bibitem[{Shay {et~al.}(2007)Shay, Drake, \& Swisdak}]{shay07a}
Shay, M.~A., Drake, J.~F., \& Swisdak, M. 2007, Phys. Rev. Lett., 99

\bibitem[{Somov \& Kosugi(1997)}]{somov97a}
Somov, B.~V., \& Kosugi, T. 1997, Ap. J., 485, 859

\bibitem[{Wang {et~al.}(2001)Wang, Wu, Wang, Chao, Lin, \& Yoon}]{wang01a}
Wang, X.~Y., Wu, C.~S., Wang, S., Chao, J.~K., Lin, Y., \& Yoon, P.~H. 2001,
  Ap. J., 547, 1159

\bibitem[{Wu(1996)}]{wu96a}
Wu, C.~S. 1996, Ap. J., 472, 818

\bibitem[{Zeiler {et~al.}(2002)Zeiler, Biskamp, Drake, Rogers, Shay, \&
  Scholer}]{zeiler02a}
Zeiler, A., Biskamp, D., Drake, J.~F., Rogers, B.~N., Shay, M.~A., \& Scholer,
  M. 2002, J. Geophys. Res., 107, 1230

\end{thebibliography}

\clearpage

\clearpage

\clearpage

\clearpage

\end{document}